\documentclass{ws-procs975x65}

\begin{document}



\title{HEAD-ON COLLISIONS OF DIFFERENT INITIAL DATA}

\author{ULRICH SPERHAKE, BERND BR\"UGMANN, JOS{\'E} A.~GONZ{\'A}LEZ, \\
MARK D.~HANNAM, SASCHA HUSA
}

\address{Theoretical Physics Institute,\\
University of Jena, 
D-07743 Jena, Germany 
}


\begin{abstract}
We discuss possible origins for discrepancies observed in the radiated
energies in head-on collisions of non-spinning binaries starting from
Brill-Lindquist and superposed Kerr-Schild data. For this purpose,
we discuss the impact of different choices of gauge parameters and
a small initial boost of the black holes.

\end{abstract}

\bodymatter

\section{Introduction}\label{intro}
The area of numerical relativity has made dramatic progress in the last
two years
\cite{Pretorius2005a, Campanelli2006, Baker2006}
and numerical simulations
of black hole binaries performed by various groups have resulted in
a wealth of literature on astrophysical topics and those related to
gravitational wave data analysis.
At the same time, laser-interferometric GW detectors, LIGO, GEO600, TAMA,
VIRGO, have started collecting data at design sensitivity. The area
of GW physics has thus entered a very exciting stage with vast potential
for astrophysics, our understanding of the early universe and fundamental
physics.

From the viewpoint of numerical relativity, though,
a number of important questions still remain to be addressed. These
largely concern the accuracy of the produced waveforms, the dependency
of the results on numerical techniques, their detailed matching with
results predicted by approximation theories as well as the mass production
of waveforms covering the complete parameter space for use in
GW observations and parameter estimation.

The purpose of this study is to address the dependency of the numerical
results on the choice of black-hole binary initial data. In contrast
to spacetimes containing single stationary black holes, there exist
no uniqueness theorems guaranteeing that two data sets for binary black holes
using different data types represent the same physical configuration. Indeed,
such data sets are known to generally differ in the amount of gravitational
radiation inherent in the initial data.

The dependency on initial data parameters (though not data type)
has been studied in the case of binary black hole coalescence
in Refs.~\citen{Baker2006a, Campanelli2006b, Bruegmann2006a}.
Using the moving puncture
technique, the merger waveforms are found to agree well for different
initial separations and algorithms to produce quasi-circular initial
configurations.
A comparison of GWs
produced in the evolution of Cook-Pfeiffer and puncture data
using different evolution techniques has been presented in
Ref.~\citen{Baker2007a} and shown good agreement. There remains a
difficulty in the identification of free initial parameters in this case,
however
(cf.~the non-vanishing spin in the Cook-Pfeiffer data set in this
comparison).
This identification of parameters represents a simpler and cleaner
task in the case of head-on collisions of non-spinning black holes
which has been studied in Ref.~\citen{Sperhake2006}. That study observed
systematically larger amplitudes by about $10~\%$
in the merger waveform resulting
from Kerr-Schild data compared with those of Brill-Lindquist data.
Here we investigate two possible causes for this discrepancy:
the dependency of the results
on the gauge trajectories in the case of Kerr-Schild data and the impact
of deviations from time symmetry of the initial data.

\section{Results}

The simulations presented in this work have been obtained with the
{\sc Lean} code \cite{Gonzalez2007a, Sperhake2006}
which uses the BSSN formulation
of the Einstein equations together with the moving puncture approach
\cite{Campanelli2006, Baker2006}.
It is based on the {\sc Cactus} \cite{Goodale2002}
computational
toolkit and the {\sc Carpet} \cite{Schnetter2004} mesh-refinement package.
For a detailed description of the code as well as the construction of
initial data we refer the reader to Ref.~\citen{Sperhake2006}.
%
\begin{table}[b]       

\tbl{Parameters for the black hole models.
     The gauge parameters in columns 5 to 9 are only used
     for the Kerr-Schild simulations. There we also use
     $q_0^z=-0.000278$, $-0.000165$ and $-0.000104~M^3$
     respectively for models 2a, 3 and 4. Here $M$ is the ADM mass
     of the system.
     The two rightmost columns list the energy radiated in the $\ell=2$,
     $m=0$ mode for both data types,
     ignoring contributions due to the spurious initial burst.}
{\begin{tabular}{@{}cccc|ccccc|cc@{}}
\hline\\
&&&&&\\[-15pt]
Model & $\frac{D_{\rm KS}}{M}$ & $\frac{D_{\rm BL}}{M}$ & $\%\frac{E_{\rm b}}{M}$
      & $v_0^z$ & $a_0^z~M$ & $j_0^z~M^2$ & $\frac{t_1}{M}$ & $\frac{t_2}{M}$
      & $\%\frac{E_{\rm KS}}{M}$ & $\%\frac{E_{\rm BL}}{M}$ \\
\hline\\
&&&&&\\[-15pt]
1a & 10 & 8.6  & 2.8 & 0 & -0.037 & 0.0038 & 10 & 35 & 0.066 & 0.051 \\
1b & 10 &      & 2.8 & -0.08 & -0.0061 & -0.0002 & 20 & 40 & 0.065 & \\
2a & 12 & 10.2 & 2.4 & 0 & -0.029 & 0.0040 &  25 & 50 & 0.067 & 0.052 \\
2b &    & 10.2 & 2.4 & \multicolumn{5}{l|}{(with initial physical boost $v=0.067$)}     & & 0.0525\\
3 & 14 & 12.5 & 2.0 & 0 & -0.022 & 0.0027 &  25 & 57 & 0.067 & 0.052 \\
4 & 16 & 14.6 & 1.6 & 0 & -0.018 & 0.0020 &  34.5& 84.7 & 0.086 & 0.054 \\
\hline
\end{tabular} \label{tab: models}}
\end{table}
%

We first discuss the gauge trajectories used in Ref.~\citen{Sperhake2006}
for the Kerr-Schild data.
There, algebraic gauge conditions are constructed which
require trajectories for the (approximate) black hole positions
(see \citen{Sperhake2006} for details). These are prescribed as polynomials
$\pm x^i(t)=x_0^i + v_0^i t + a^i t^2/2 + j^i t^3/6 + q^i t^4/24$ which are
smoothly (up to fourth derivatives) matched to the static function
$x^i(t)=0$ in a time interval $t_1 < t < t_2$. Here
$x_0^i$, $v_0^i$, $a_0^i$, $q_0^i$, $t_1$ and $t_2$ are free parameters
which need to be chosen carefully to avoid numerical instabilities. In
Table \ref{tab: models} we list the values for each (Kerr-Schild) model.
In order to assess the impact of the particular choice of these parameters,
we have evolved the initial data of model 1 with alternative gauge parameters
as listed in the second row of the table. This alternative gauge trajectory
is motivated by the initial coordinate velocity $v=-0.08$ of the
central position of the apparent horizon as measured
using Thornburg's {\tt AHFinderDirect} \cite{Thornburg1996, Thornburg2003}.

\begin{figure}
\begin{center}
  \includegraphics[angle=-90,width=145pt]{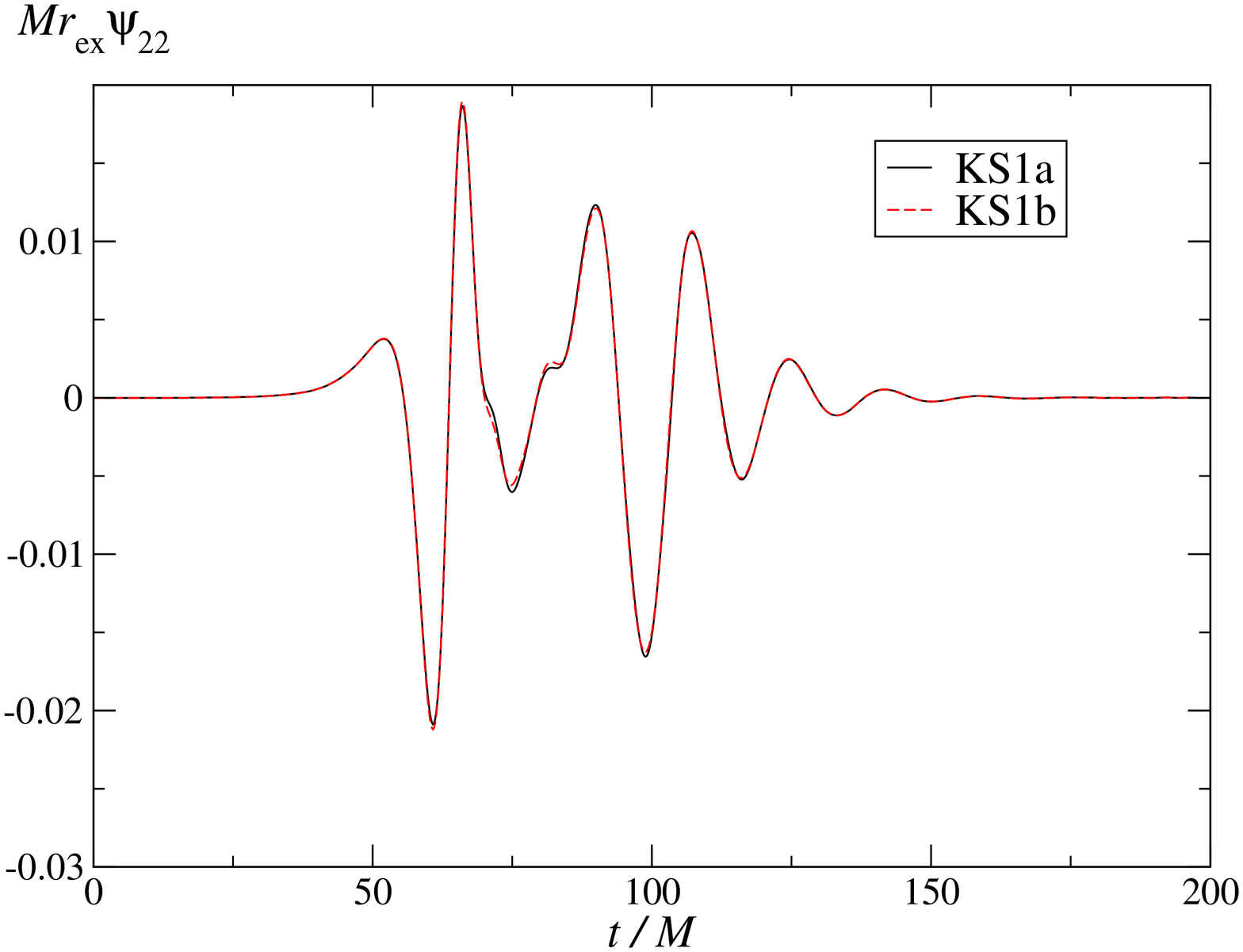}
  \includegraphics[angle=-90,width=145pt]{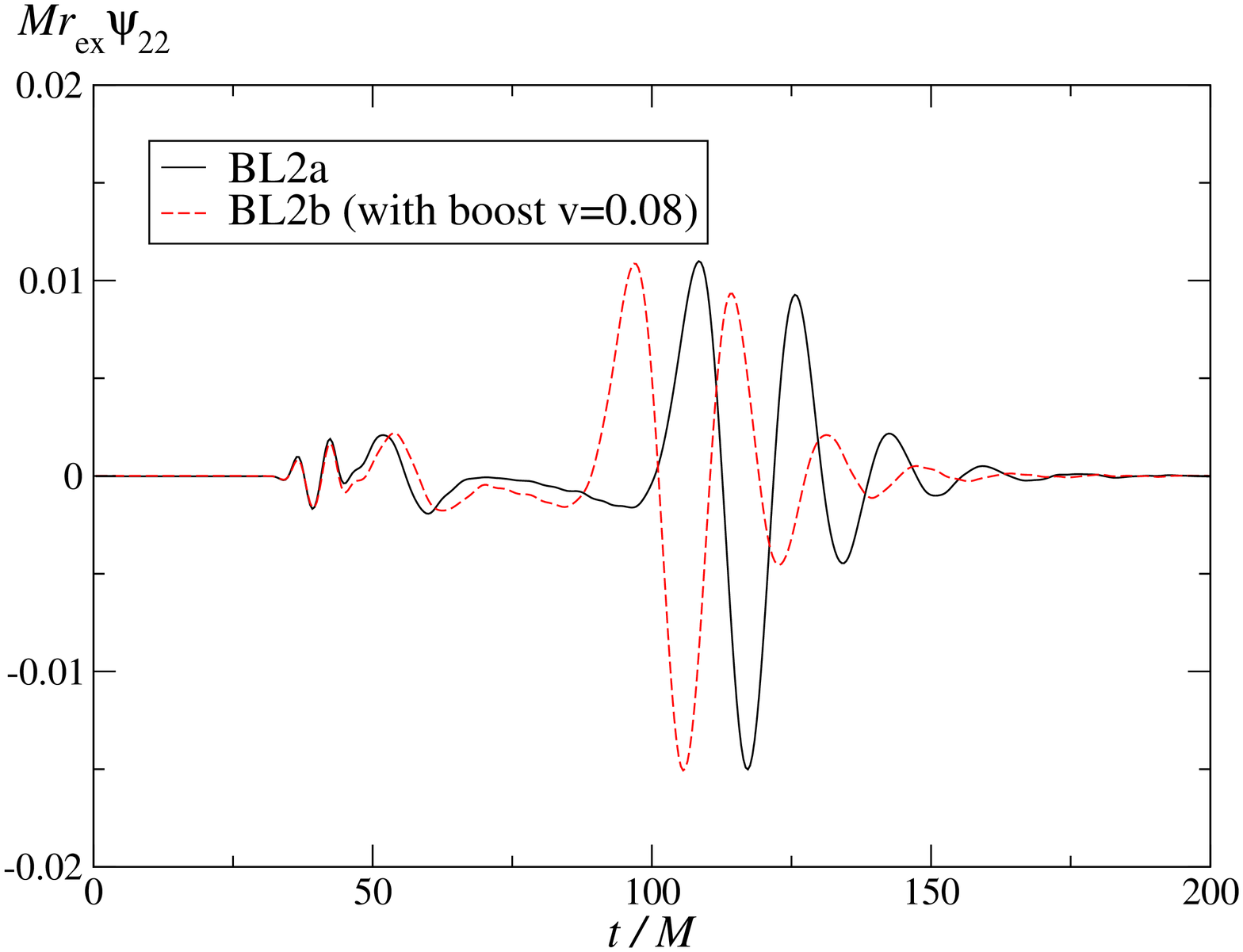}
  \caption{The $\ell=2$, $m=0$ mode of the scaled Newman Penrose scalar
           $Mr_{\rm ex}\psi_4$ extracted at $r_{\rm ex}=50\,M$ for models
           1a, 1b (left panel) and models 2a, 2b (right panel).}
  \label{fig: waves}
\end{center}
\end{figure}

The resulting waveforms are shown
in the left panel of Fig.~\ref{fig: waves}.
Both the waveforms and the radiated energies thus obtained for
model 1 show excellent agreement. The
differences in radiated energy are about $1.5~\%$  and thus
substantially smaller than
the discrepancies between $E_{\rm KS}$ and $E_{\rm BL}$.

Second, we assess the impact of deviations from exact time symmetry of the
initial superposed Kerr-Schild data. These deviations manifest themselves
in a small but non-vanishing initial coordinate velocity of the superposed
Kerr-Schild holes as measured by the central position of the apparent horizon.
For the case of model 2 we have measured this velocity
to be $v=0.067$.
In order to estimate what impact such an initial velocity has on the
resulting waveforms, we have applied an initial linear momentum $p_z=m~v$
to the Brill-Lindquist version of this model, where $m$ is the irreducible mass
of a single hole. The resulting waveform
is compared with its non-boosted counterpart in the right panel
of Fig.~\ref{fig: waves}.
Again, the wave amplitudes show good agreement, as do
the resulting values for the radiated energy in Table \ref{tab: models}.

In summary, we find the observed differences in radiated energy
resulting from modifications of the gauge trajectories and a possible
initial boost of the black holes to be of the order of $1~\%$ and thus
substantially below the differences of about $>20~\%$ observed in
Ref.~\citen{Sperhake2006} between the two types of initial data.

\vspace{-0.35cm}

\section*{Acknowledgments}
This work was supported by
DFG grant SFB/Transregio~7 ``Gravitational Wave Astronomy'',
and the DEISA Consortium (co-funded by the EU, FP6 project
508830).
computations were performed at LRZ Munich and HLRS, Stuttgart.

\vspace{-0.35cm}


%
%


\end{document}